\documentclass[12pt]{iopart}

\usepackage{graphicx}
\usepackage{lineno}
\begin{document}

\title{Oscillations arising when switching off a discontinuous magnetic field}

\author{P M Romanets}

\vspace{10pt}
\begin{indented}
\item[]August 2019
\end{indented}

\begin{abstract}
In the present paper, we theoretically study the effect of density of state variation on the phenomena in the discontinuous magnetic field. Special attention is paid to the transient processes when the magnetic field is switched on (off). The difference in the density of states for the free 2D electron gas and the electron gas on under the magnetic field can lead to the occurrence of the electrostatic potential. Considering the transient phenomena of magnetic field switching, we obtain oscillating behavior of the electrostatic potential.   The analysis includes two limit cases when the oscillation frequency of the electrostatic potential is low relative to the cyclotron frequency and when they are of the same order.
\end{abstract}

%
%
%
%
%

\section{Introduction}

A new wave of interest in a highly-heterogeneous magnetic field took place about twenty years ago. A large number of interesting effects were investigated experimentally and theoretically. Magnetic antidots, quantum dynamics of electrons under the action of a discontinuous magnetic field in the presence of a background magnetic field, magnetic superlattice, diffusion in a discontinuous magnetic field, and many other interesting problems were investigated \cite{19,1,20,21}. However, to the best of our knowledge, the effects associated with varying the density of states near the magnetic field step are not considered. Moreover, as will be shown below, some of the theoretical results obtained before should be corrected or refined, since they did not take into account the effects of the density of states variation.

On the other hand, there is still a strong interest in researching processes on the rapid dynamics of the magnetic field, which have obvious prospects for practical application. Practically, these studies have two directions: the improvement of methods of ultrafast magnetization of ferromagnetic materials and the improvement of measurement methods.
The rate of magnetization of a ferromagnet depends on the chosen method and on the structural features of the material itself. The rapid dynamics of the magnetic field can be achieved with the help of micromagnets \cite{7}, but the best results were achieved in experiments on the use of a femtosecond laser \cite{5}.
With the help of a femtosecond laser, it is possible not only to demagnetization but also to magnetizing ferromagnets. For this use a static magnetic field \cite{8}, or circularly polarized light \cite{6}. In such experiments, the rate of reversal is limited mainly by the structural features of the ferromagnet. Some influence of temperature and geometry of the sample is also possible \cite{18}.
The time scale of demagnetization for the ferromagnets from the transition metal group \cite{7,5,8,6,9}, groups of rare earth metals \cite{16,17} and ferromagnetic insulators differ significantly \cite{10}. For the transition metal group, the demagnetization time with the femtosecond pump is the smallest and is within the range of 50-250 fs, for rare earth metals, the order is 1ps, whereas for ferromagnetic insulators the time is measured by tens or hundreds of peak seconds.
Among the most popular detecting methods is the magneto-optical Kerr effect (MOKE) and the x-ray Magneto-Circular Dichronism (XMCD). MOKE has a better time resolution, but all structural features of the magnetization of a ferromagnetic compound are averaged spatially \cite{9,11,13,15,2,14}. XMCD, on the contrary, has a worse time resolution but can reflect the structural features of a ferromagnetic compound \cite{9}. Note that all measurement methods are optical. We hope that the following results will be useful for the development of non-optical measurement methods for processes involving the rapid dynamics of the magnetic field.

\section{The field caused by the density of states variation}
In this section, we consider the field caused by the difference in the density of states for the free two-dimensional electron gas (2DEG) and the 2DEG under the magnetic field. Figure 1(a) shows the experimental scheme. Due to the Meissner effect, the magnetic field in a quantum well (QW) is sharply dependent on the coordinate $B(y, t) = B_0(t)\theta(y)$. Below, everywhere we assume that the dependence of the magnetic field on time has form $B_0 (t) = B_m [1-\exp(-\gamma t)]$ and $B_0 (t) = B_m \exp (-\gamma t)$ when switching on and off correspondingly.
\begin{figure}
	\centering
	\includegraphics[width=80mm]{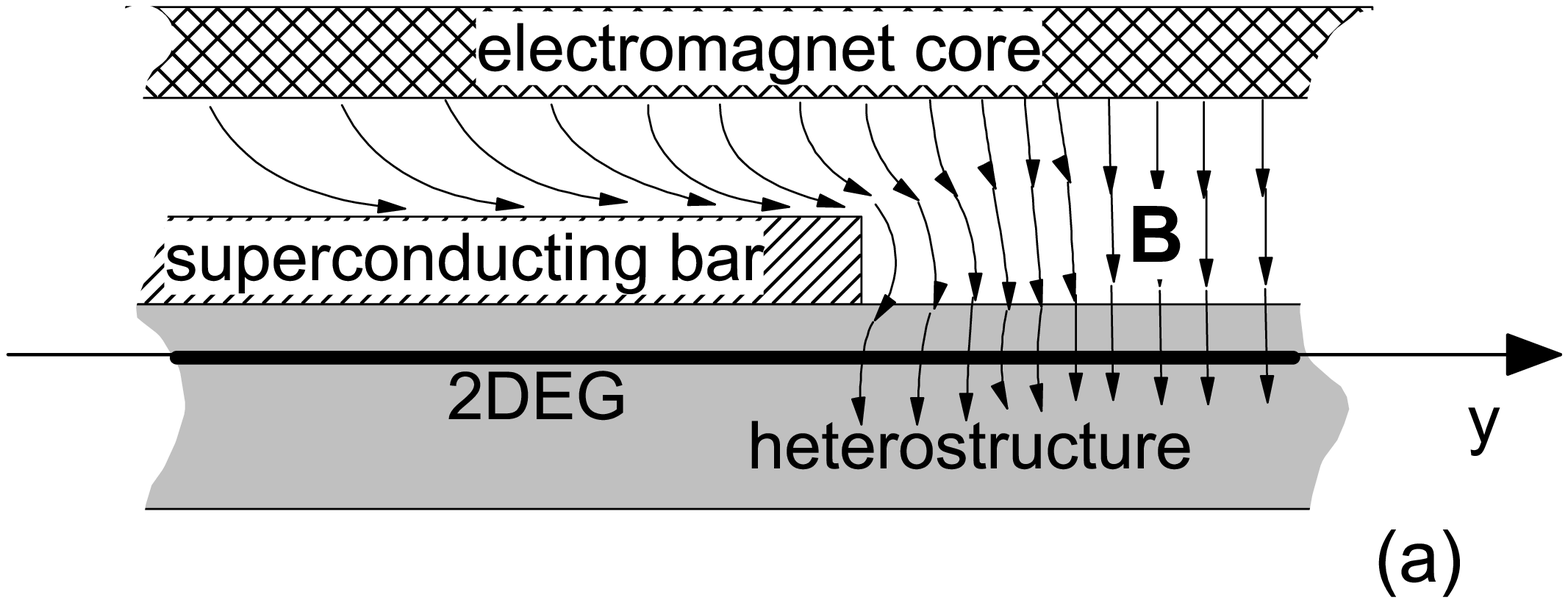}
		\includegraphics[width=60mm]{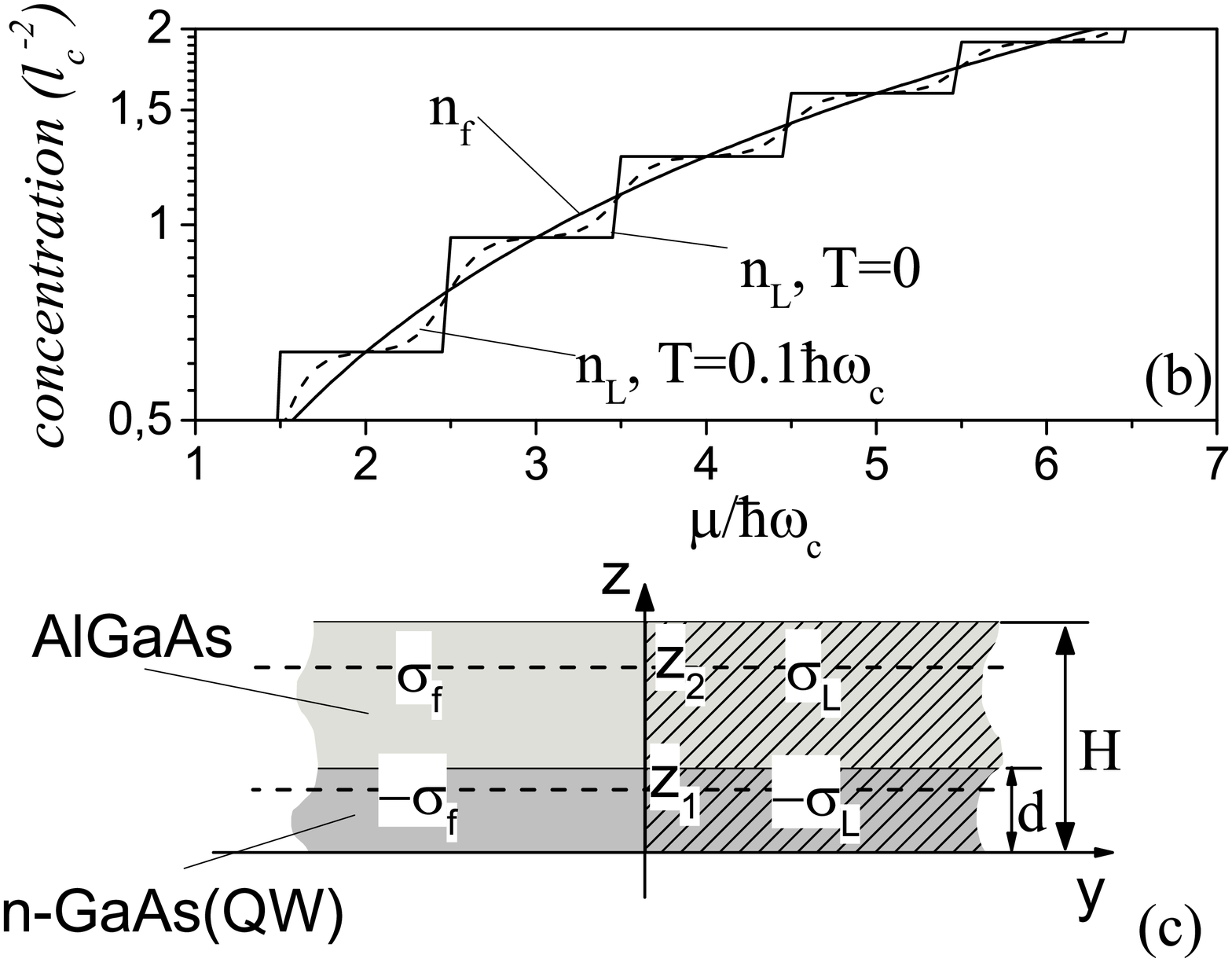}
	\caption{Frame (a) depicts the experimental schema. Frame (b) shows the plot in half-logarithmic scale with concentration of free electrons $n_f$ ($y<0$) and the localized on Landau levels electrons $n_L$ ($y>0$). The schema on the frame (c) explains how potential is calculated (see Appendix A); $\sigma_{f,L}$ denotes the corresponding surface density of charges.}
\end{figure}
 For the cyclotron frequency $\omega_c (t) = \omega_c^m [1-\exp(-\gamma t)]$ 
and $\omega_c (t) = \omega_c^m \exp (-\gamma t)$ respectively.
\begin{eqnarray}\label{concentrations}
n_f (\mu) \simeq \frac{mT}{\pi\hbar^2} \ln [1 + \exp (\mu / T)],\nonumber
\\
\left.n_L (\mu, t)  \right|_ {\{\tau_n \to \infty\}} \simeq \frac 1 {\pi l_c^2 } \sum_{n = 0}^{\infty} \left\{ 1 + \exp  \left[\omega_c(t) \hbar \left(n + \frac 12\right) / T  -\mu / T\right]\right\}^{-1},
\end{eqnarray}
here $\mu$ is the chemical potential, $T$ is the temperature, and $n_f$ and $n_L$ are the concentrations of free electrons and electrons at the Landau levels, $\tau_n$ are the quatum life-times for $n$-th Landau level. On Figure 2(b) one can see the dependence of the concentrations on the chemical potential for temperatures $T = 0$ and $T = 0.1$  $\omega_c\hbar$. Obviously, the concentration $n_f$ equals $n_L$ only when $\mu=N\hbar\omega_c/2$ ($N$ is a positive integer), or when $T \geq\hbar\omega_c$. The difference in concentration will lead to the emergence of an electric field. The heterostructure as a whole should be electroneutral. In the assumption of a QW rectangular profile, the charge density for a heterostructure can be written as follows:
\begin{eqnarray}\label{charge_density}
\rho(y, z) \simeq \left\{
\begin{array}{lll}
\frac{-e\Delta n\sin^2( z
	\frac{\pi}{d})}{d},\qquad 0<z<d, y<0;\\

\frac{e\Delta n\sin^2( z\frac{\pi}{d})}{d},\qquad 0<z<d, y\geq0;\\
\frac{e\Delta n}{2H},\qquad d\leq z\leq H, y<0\\
\frac{e \Delta n}{2H},\qquad d\leq z\leq H, y\geq0\\

\end{array} 
\right.
\end{eqnarray}
where $\Delta n = n_f-n_L$,  $e$ is the elementary charge, $d$ is the QW width, $H$ is the distance from the lower edge of the QW to the interface of the heterostructure (see Figure 2 (b)). 

Because of the non-uniform distribution of charge (\ref{charge_density}) in QW, an electric field is generated. The statical screening effects will be considered below. The dynamical screening will not be taken into account.The potential of this field can be represented as (see Appendix A):%
\begin{eqnarray}\label{eq_potential}
\phi(y,t)=\frac{2\pi e}{\epsilon}\Delta n(t)\left[\left(\frac{d-H}{8}-\frac{3d}{16\pi^2}\right)\right.\nonumber\\\left. + \sum_{n=1}^{\infty}{a_n}\exp(-|y\lambda_n|)
\right]sgn(y)
\end{eqnarray}	
where the coefficients defined with the equations:
\begin{eqnarray}\label{eq_potential_koef1}
a_n= -u_n\frac{4\sin(\lambda_nd)}{\left[\lambda_nd(4-\lambda_n^2d^2)\right]}
\end{eqnarray}	
and
\begin{eqnarray}\label{eq_potential_koef2}
u_{n\ne0} = (-1)^n\frac{\left\{\sin\left(\frac{2\pi H}{d}\right)d^3\lambda_n^2/H-2\pi\lambda_n^2d^2+8\pi^3\right\}}{\pi d\lambda_n^2\left(4\pi^2-\lambda_n^2d^2\right)};\nonumber\\
u_{n=0} = \frac{4H^3\pi^3-6\pi^3dH^2+3\sin\left(\frac{2\pi H}d\right)-6d^2\pi H}{24Hd\pi^3}.
\end{eqnarray}
When the slow dynamics of magnetic filed takes place the statical screening is essential. It could be taken into account as follows:
\begin{eqnarray}\label{stat_screen}
\Delta n(t) \simeq n_L(\mu,t)-n_f(\mu)+\nonumber\\\sum_{k=0}^2\left[\frac{\partial n_L(\mu,t)}{\partial \mu}\left.\frac {e^k\phi(y,t)^k}{k!}\right|_{y\to-\infty}-(-1)^k\frac{\partial n_f(\mu)}{\partial \mu}\left.\frac {e^k\phi(y,t)^k}{k!}\right|_{y\to\infty}\right]
\end{eqnarray}
where we neglect attenuation of the static screening in the small region $|y| \lambda_1^{-1}$. Taking into account formulas (\ref{concentrations}) and (\ref{eq_potential})-(\ref{eq_potential_koef2}) one could obtain polynomial equation relative to $\Delta n(t)$. The series cut off is justified by a numerical estimation.
In Fig. 2 (a) depicts the coordinate dependence of the electrostatic potential. The used approach supposes that potential is asymmetric. The different lines correspond to different chemical potentials. Note, that maximum value of the potential corresponds to $\mu=9.2\hbar\omega_c^^m$ when $T=0.1\hbar\omega_c^m$, whereas in the low temperature limit $T\to0$, the maximum value corresponds $\mu=9.5\hbar\omega_c^^m-0$. The effect of temperature is clearly shown in Fig. 2(b) and also could be understood from Fig. 1(b). The temporal oscillations of the potential appearing when the magnetic field is switching on/off are described in Fig 2(c).
The electrostatic potential $\phi(y, t)$ is the function of time at the point $y=H/2$. The  used in calculations temperature equals $0.1\times\hbar\omega_c^m$.
It is seen that there are arbitrarily high frequencies, but the amplitude decreases as the frequency increases. The frequencies could be estimated with the next formulas
\begin{eqnarray}\label{eq_min_fr}
\Omega_{on}(\omega_c)=\frac{-2\pi\gamma}{\ln\left[\frac{\frac{\mu}{\hbar\omega_c}-\frac{\mu}{\hbar\omega_c^m}-1}{(\frac{\mu}{\hbar\omega_c}-1)(\frac{\omega_c}{\omega^m_c}-1)}\right]},\quad\Omega_{off}(\omega_c)=\frac{-2\pi\gamma}{\ln\left[1-\frac{\hbar\omega_c}{\mu}\right]},
\end{eqnarray}
The minimum frequencies are realized at the  maximum amplitude (see Figure 3), whereas for $\lim_{\omega_c\to\omega_c^m}(\Omega_{on})=\infty$ and $\lim_{\omega_c\to0}(\Omega_{off})=\infty$ corresponding amplitudas are zero.

\begin{figure}
	\centering
	\includegraphics[width=80mm]{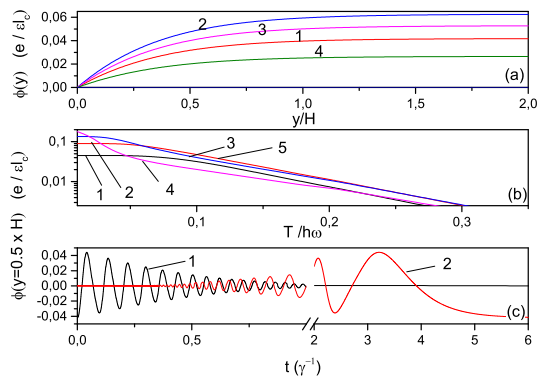}
	\caption{The properties of the screened electrostatic potential in the quasistationary case ($\gamma<<<\omega_c$). Frame (a) shows the coordinate dependence of the potential for different chemical potentials $\mu=9.1,~9.2,~9.3,~9.4\times\hbar\omega_c$ (lines 1-4 correspondingly) and for the temperature $T=0.1\hbar\omega_c^m$. The frame (b) shows the temperature dependence of the potential. The lines numbering is the same as on the frame (a). The frame (c) shows the time-dependence of the potential $\phi$ at the point $y=H/2$. The chemical potential and the maximum cyclotron frequency related as $\mu=9.5\hbar\omega_c^m$. The oscillations arising when switching off (line 1) and on (line 2) of the magnetic field at the time moment $t=0$. The time in units $\gamma^{-1}$. The heterostructure geometry is defined by the next parameters $H=5.8l^{min}_c$ and $d=0.2 l^{min}_c$. The influence of relaxation processes is neglected.}
\end{figure}
\section{Bondary ground state}
Equations (\ref{charge_density})-(\ref{eq_potential_koef2}) do not take into account the tails of states localized in the region $y\in(0, l_c)$. Thus states should be treated as boundary states because they essentially penetrate the magnetic field-free region.  In fact, in \cite{1}, the authors analyzed in detail the boundary states that arise at the boundary of the magnetic field breaking.  In particular, the case corresponding to $B(x, t) = B_m \theta(y)$ was analyzed. In the Schroedieuer equation, the authors did not include the potential of type $\phi(y, t)$, so if $T\ll\hbar\omega_c$ their results are true only in the case  $\mu=\hbar\omega_c(N_{LL}+\frac12)$, where $N_{LL}$ is the positive integer. Specifically when $\hbar\omega_cN_{LL}<\mu<\hbar\omega_c(N_{LL}+\frac12)$ the concentration of electrons in Landau levels is lower than that for the free electrons and the boundary states could arise. Indeed, the superposition of the effective potential for the magnetic field and the potential $\phi(y,t)$ can lead to the electron confinement (see Fig. 2a, taking into account the change of sign for the potential energy of an electron due to a negative charge). 
We have used approach of variatinal method with trial wave-function
 \begin{eqnarray}\label{charge_density}
 \Psi_0(x, y) \simeq \left\{
 \begin{array}{lll}
 A_>\exp(-(y-y_0)^2/a^2), \qquad y>0;\\
 
 A_{\le}\exp(-(2y_0y-y_0)^2/a^2), \qquad y\le 0; 
 \end{array} 
 \right.
 \end{eqnarray}
 The results of the computation are shown in Fig. 3. One can see that the deeper states appear on higher distance from the boundary $y0$, on the other hand, they are more sensitive to the temperature. Particularly, ground state for $y0=0.3l_c$ disappears at $T\ge0.32\hbar\omega_c$, whereas states for $y0=0.1l_c$ disappears only when $T\ge0.35\hbar\omega_c$ (compare Fig. 3 (c) and (d)). n the limit case $y0\to\infty$ we should obtain Landau solution for the ground state, i.e. $a=\sqrt{2}l_c$. Comparing this value with the values in Fig. 3 (a,b) we can conclude that the boundary ground states demonstrate higher localization than the corresponding Landau states.
 \begin{figure}
 	\centering
 	\includegraphics[width=80mm]{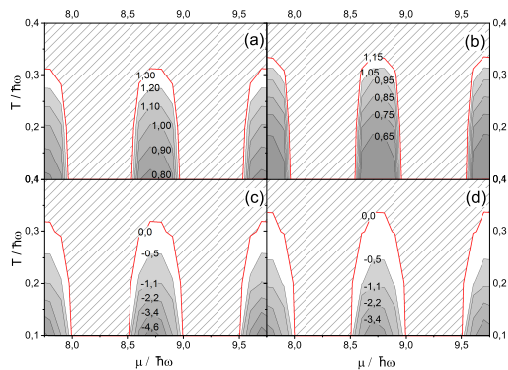}
 	\caption{Contour plots on frames (a) and (b) depict the parameter $a$ obtained from the variational method versus temperature and chemical potential. The corresponding ground state energies are shown on frames (c) and (d). The case on frames (a) and (c) is for $y_0=0.3 l_c$ whereas (b) and (d) is for $y_0=0.1l_c$. }
 \end{figure}
\begin{figure}
	\centering
	\includegraphics[width=80mm]{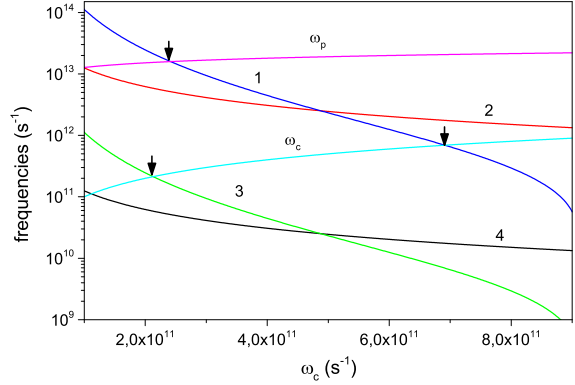}
	\caption{The cyclotron and magnetoplasmon resonance condition. The lines 1, 2 describe $\Omega_{on}$, $\Omega_{off}$ when $\gamma=2\times10^{10}$ $s^{-1}$ and lines 3, 4 are the same for $\gamma=2\times10^8$  $s^{-1}$ vs. cyclotron frequency. The short arrows are the markers for resonant conditions. The cyclotron and magnetoplasmon frequencies are $\omega_c$ and $\omega_p$. The conxcentration of 2D electrons is $1.8\times10^{11}$ $cm^{-2}$.}
\end{figure}
We should note, that the effect of the state density in the case of an asymmetric magnetic field $B(x, t) = B_m sgn (y)$ is not significant.

\section{Ultrafast switching of the magnetic field}
Under the slow magnetic field switching and low temperature, the relaxation processes are not significant. On the other hand, if the ultrafast switching of the magnetic field takes place, then $\Omega^{nim}_{on, off}\sim\omega^m_c$. Under these conditions, the quantum lifetimes on Landau levels $\tau_n$ are essential parameters.
The faster switching of the magnetic field the shorter quantum lifetime at the certain Landau level. In the limit case, one obtains that Landau levels are an incorrect approach. In this section, we consider the influence of the finite quantum lifetime on the phenomena described in the first section. Nevertheless, below we suppose $\omega^m_c\gg\gamma$ because of using of the perturbation theory. As well one must modify the functions describing the time-dependent magnetic field. After transformation, the cyclotron frequency has the next form:
\begin{eqnarray}\label{modified_wc}
\omega_c(t)=\left\{
\begin{array}{lll}
\omega_c^m\exp[-\gamma \int_{-\infty}^tdt'f_{\beta}(t')]+\omega_c^b,\quad \mbox{switch~off};\nonumber\\
\omega_c^m\left\{1-\exp[-\gamma \int_{-\infty}^tdt'f_{\beta}(t')]\right\}+\omega_c^b,\quad \mbox{switch~on};
\end{array}
\right.
\end{eqnarray}
where $\omega_c^b\ll\omega_c^m$ is due to small background magnetic field and $f_{\beta}(t)$ is any smooth function satisfying the condition $\lim_{\beta\to \infty} [f_{\beta}(t)]=\theta(t)$, besides $\gamma,\beta\ll\omega_c^m$. This means that the wave function $\Psi_n(t,x,y)$ is obtained by the substitution $\omega_c = \omega_c(t)$ in the stationary solution of the Schrodinger equation $\Psi_n^{(st.)}(x, y)$  is a good approach (see \cite{LandauLifshitzIII} p. 424, the reduction to the 2D case is obvious):
\begin{eqnarray}\label{zero_appr_Wave_funct}
\Psi_n(t,y,z) \simeq \exp\left[\int_0^t\hat{H}_0(t')dt'\right]\left.\Psi^{st.}_n(y,z)\right|_{\omega_c=\omega_c(t)}
\end{eqnarray}
The time-dependent density of states could be introduced as follows:
\begin{eqnarray}\label{density_of_st0}
\rho_{\gamma}(E,t)=\nonumber\\\frac{1}{2\pi}\sum_{n=0}^{\infty}\int_{-\infty}^0\exp\left(i\frac{E}{\hbar}\Delta t+\int_{t-\Delta t}^{t}\frac{dt'}{\tau_{n}(t')}\right)\left<\Psi_n,t|\Psi_n,t+\Delta t\right> d\Delta t + c.c.;
\end{eqnarray}
The distribution function $f_{\gamma}(E,t)$ and the density of states must satisfy 
\begin{eqnarray}\label{density_vs_distr}
\rho_{\gamma\to 0}(E,t)=\frac{1}{2\pi l_c^2}\sum_{n=0}^{\infty}\delta\left[E-\hbar\omega_c\left(n+\frac 12\right)\right],\nonumber\\ f_{\gamma\to 0}(E,t)=\left[1+\exp\left(\frac{E-\mu}T\right)\right]^{-1};
\end{eqnarray}
After rewriting them in the from $\rho_{\gamma}(E,t)=\rho_{\gamma\to 0}(E,t)+\Delta\rho_{\gamma}(E,t)$ and  $f_{\gamma}(E,t)=f_{\gamma\to 0}(E,t)+\Delta f_{\gamma}(E,t)$ and neglecting $\Delta\rho_{\gamma}(E,t)\Delta f_{\gamma}(E,t)$ one could calculate the concentration in the region $y>0$ as $n_L(t)=\int_0^{\infty}\rho_{\gamma}(E,t)f_{\gamma\to 0}(E,t)+\rho_{\gamma\to 0}(E,t)\Delta f_{\gamma}(E,t)dE$
where the second term is zero because of (\ref{B10}). Transforming the hamiltonian $\hat{H}_0(t')$ to act locally in time one could obtain
\begin{eqnarray}\label{density_of_st1}
\rho_{\gamma}(E,t,\bar{y})=\frac 1{2\pi l_c(t)^2}\sum_{n=0}^{\infty}\int_{-\infty}^0\exp\left[i\frac{E}{\hbar}\Delta t-i\varphi(t,t+\Delta t)\left(n+\frac 12\right)+\int_{t-\Delta t}^{t}\frac{dt'}{\tau_{n}(t',\bar{y})}\right]\nonumber\\\times A_n(t,t+\Delta t,\bar{y}) d\Delta t+c.c.,
\end{eqnarray}
where
\begin{eqnarray}\label{phase}
 \varphi(t,t')=\frac 12\left[\omega_c(t')(t-t')+\frac 1 {\omega_c(t')}\int_{t'}^tdt''\omega_c^2(t'')\right],\nonumber\\
A_n(t,t',\bar{y})\simeq\frac{1}{2\pi}\left(\frac{1}{2}\right)^{n-\frac 12}\left(\sqrt{\frac{\omega_c(t')}{\omega_c(t)}}+\sqrt{\frac{\omega_c(t)}{\omega_c(t')}}\right)^{n-\frac 12}\nonumber\\\times \exp\left\{-\frac{\bar{y}^2m}{8\hbar}\frac{[\omega_c(t)^2-\omega_c(t')^2]^2}{\omega_c(t)\omega_c(t')^2+\omega_c(t')\omega_c(t)^2}\right\};
\end{eqnarray}
whereas $\tau_{n}(t',\bar{y})$ is defined by (\ref{B11}). The equation (\ref{density_of_st1}) demonstrates two peculariries because of the ultrafast switching of the magnetic field. The first one is because of retarding effect. Namely, the energy of the $n$-th level is not definited by current cyclotron frequancy $\omega_c(t)$, but by some time-averaged cyclotron frequancy (see $\varphi$ in \ref{phase}). We neglact this effect  below, supposing that at least one of the conditions $\gamma\ll\omega_c(t)$ or $\gamma\ll\tau_n^{-1}(\bar{y},t)$ is satisfied. The second pecularity is because of finite quantum life-time. We will use local in time approach $A_n(t,t',\bar{y})\approx1$, $ \varphi(t,t')=\omega_c(t)(t-t')$ and $\int_{t-\Delta t}^{t}\frac{dt'}{\tau_{n}(t',\bar{y})}\approx\frac{\Delta t}{\tau_{n}(t,\bar{y})}$ to study it influence on the electrostatic potential. Then the additional electrons concentration under the ultrafast switching magnetic filed and $T\to0$ is 
\begin{eqnarray}\label{qa_dconcentraion}
\Delta\tilde{n}(y,t) \simeq\frac 1{\pi^2 l_c(t)^2}\\\nonumber\times\sum_{n=0}^{\infty}\left\{\arctan\left[\frac{\mu/\hbar-\omega_c(n+\frac 12)}{\tau^{-1}_{n}(t,y)}\right]+\arctan\left[\frac{\omega_c(n+\frac 12)}{\tau^{-1}_{n}(t,y)}\right]\right\} - n_f(\mu);
\end{eqnarray}
\begin{figure}
\centering
\includegraphics[width=80mm]{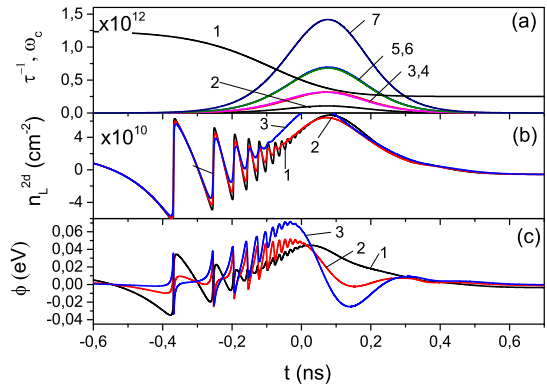}
\caption{A finite quantum life times modify the oscillation. The frame (a) shows cyclotron frequancy (line 1) and invese quantume life times for the different quantum levels $n=10,~20$ and different distance from the magtentic field step $y$; the line 2 for $n=10$ and $y=0.1$ $\mu m$, the line 3 for $n=10$ and $y=0.2$, the line 4 for $n=20$ and $y=0.1$ $\mu m$, the line 5 for $n=10$ and $y=0.3$ $\mu m$,the line 6 for $n=20$ and $y=0.2$ $\mu m$, the line 7 for $n=20$ and $y=0.3$ $\mu m$. The frame (b) shows the oscillations of the concentration of the electon gas under the magnetic filed $y>0$ relatively to the free electon gas concentraion; the line 1 for $y=0.1$ $\mu m$, the line 2 for $y=0.2$ $\mu m$ the line 3 for $y=0.3$ $\mu m$. The frame (c) shows electrostatic potential; the lines 1, 2 and 3 have the same meaning as on frame (b).  Calculation are performed for GaAs QW with $H=300~nm$ and $d=10~nm$, $\omega_c^m=10^{12}~s^{-1}$, $\gamma=0.08~\omega_c^m$, $\beta=0.04~\omega_c^m$, $\omega_c^b=0.25~\omega_c^m$, $\mu=9.8~ \hbar \omega_c^m$ and $T\to0$. The case of the switching off the discontineous magnetic field.}
\end{figure}
where we are not considering dynamical screening and cyclotron resonance (see Fig.4). In contrast to the eq.(\ref{stat_screen}), the statical screening is impossible in the case of ultrafast switching of the magnetic field.
The corresponding potential can be calculated using the procedure described in the last part of Appendix A. The results of the calculations described in Fig.5. In Fig. 5(a) one could see quantum lifetimes depend on both as coordinate and time. The maximal values for inverse quantum lifetimes are realized when the magnetic fields almost switched off. The inverse quantum lifetimes also rapidly increase with the coordinate $y$. In Fig 5(b) one can see that the oscillation of the concentration are quenched with the relaxation but still observable. Also, the results depict the new feature. Namely, the envelope of the time-dependent concentration increases proportionally to the inverse quantum lifetimes. This is because of quantum levels broadening. Comparing the time dependencies of the concentration and the electrostatic potential (Fig.5(c)), one could note that in the region $t\in(-0.15~ns, ~0.25~ns)$ concentration proportional to the potential derivative. To understand this feature we should take into account that in this time interval both, the potential and the concentration are strongly affected by the short lifetimes. On the other hand, the lifetimes are the functions of the coordinate. Thus the integration over the $y$ coordinate in the right-hand side of (\ref{A8}) could be transformed into the integral over the time interval.

\section{Conclusion}
We have considered the effects of density of state variation arising in a discontinuous magnetic field. It was demonstrated that in the quasi-stationary regime, the density of state variation could cause boundary states near the magnetic filed step. The result is in contradiction with the one-electron quantum calculation because the latter does not take into account average electrostatic potential arising because of the difference in concentration for the free 2D electrons and the electrons under the magnetic field.

Also, special attention was paid to the effects araising when the discontinuous magnetic field is switching on (off). The corresponding transient processes are supplemented by the oscillations of the electrostatic potential near the step of the magnetic field. Besides, we have analyzed how the phenomenon will show itself in the case of ultrafast switching off of the magnetic field when the period of densities variation is comparable to the cyclotron period. It is obtained that the finite quantum lifetimes break spatial homogeneity of the Landau states along the $y$ direction. Particularly, for  for GaAs QW with $H=300~nm$ and $d=10~nm$ and the parameters $\omega_c^m=10^{12}~s^{-1}$, $\gamma=0.08~\omega_c^m$, $\beta=0.04~\omega_c^m$, $\omega_c^b=0.25~\omega_c^m$, $\mu=9.8~ \hbar \omega_c^m$ it is expected that Landau states will be broken inside the region $y>0.3\mu m$. Whereas, the new features related to the Landau levels broadening are expected inside the region $0<y<0.3\mu m$. The brief analysis shows that oscillations could be tuned to the magnetoplasmon resonance. The detailed analysis is out of scope.
We hope that the considered above phenomena will be interesting for both, the research of discontinuous magnetic field and for ultrafast magnetization. The possible application field is to design the new methods of ultrafast magnetization detection.

\appendix
\section*{Appendix A}
\setcounter{section}{1}
To determine the electrostatic potential, the charge distribution (\ref{charge_density}), we consider the auxiliary problem [see. Fig. 2 (b)]. Let in the plane $z = z_1$ be the charges distributed with the surface density $en_f$ for $y<0$ and $en_L$ for $y>0$. Also, in the plane $z=z_2$, the surface charge density $-en_f$ for $y<0$ and $-en_L$ for $y>0$ the corresponding Poisson equation is
\begin{eqnarray}\label{A1}
 \Delta\phi_{\sigma}(y,z,z_1,z_2)=-\frac{2\pi e \Delta n}{\epsilon} sgn(y)[\delta(z-z_1)-\delta(z-z_2)]
\end{eqnarray}
we also use standart disignation for the Dirac delta function $\delta(x)$. Then, for the region $0<y<H$, the solution of the equation is
\begin{eqnarray}\label{A2}
\phi_{\sigma}(y,z,z_1,z_2)=\nonumber\\\frac{2\pi e\Delta n}{\epsilon}sgn(y)\left[\theta(z-z_1)\theta(z_2-z)\left(z-\frac{z1+z2}{2}\right)+\right.\nonumber\\\frac{z_1-z_2}{2}\left(\theta(z_1-z)-\theta(z-z_2)\right)-\nonumber\\\left.\sum_{n=1}^{\infty}\tilde{a}(z_1,z_2)\cos(\lambda_nz)\exp(-\lambda_n|y|)\right],
\end{eqnarray}
where $\lambda_n=\pi n/H$ are eigenvalues, $sgn(x)$ is the signum function and coefficients:
\begin{eqnarray}\label{A3}
\tilde{a}_{n\ne0}(z_1,z_2)=\frac{1}{2H\lambda_n^2}\sum_{j=1}^2(-1)^j\cos\left(\lambda_nz_j\right);\nonumber\\\tilde{a}_{n=0}(z_1,z_2)=\frac{z_1-z_2}{2}\left(1-\frac{z_1-z_2}{H}\right).
\end{eqnarray}
The heterostructure potential $\phi_h(y,z)$ is the superposition of the potentials $\phi_{\sigma}(y,z)$ defined by (\ref{A2})-(\ref{A3}):
\begin{eqnarray}\label{A4}
\phi_h(y,z)=\frac{2}{d(H-d)}\int_0^d\sin^2\left(\frac{\pi z_1}{d}\right)dz_1\int_d^H\phi_{\sigma}(y,z,z_1,z_2)dz_2
\end{eqnarray}
Note, that heterostructure potential satisfies the Poisson equation $\Delta \phi_h(y,z) = -4\pi\rho(y,z)/\epsilon$, where $\rho(y,z)$ in the righthand side is defined by equation (\ref{charge_density}). The potential that affects the 2D-elecrons in QW could be obtained via averaging of $\phi_h(y,z)$:
\begin{eqnarray}\label{A5}
\phi(y)=\frac{2}{d}\int_0^d\sin^2\left(\frac{\pi z}{d}\right)\phi_h(y,z)dz
\end{eqnarray}
Performing the integration, one obtains (\ref{eq_potential}). 

In the case of spatial inhomogeneouty we suppose that 2D electron concentraition $n_L$ is a smooth function of $y$ coordinate. To get potential one should replace $\Delta n$ $\to$ $\partial n^{2d}_L(y') / \partial y' \times  dy'$ and $y$ $to$ $y-y'$ in (\ref{A2}) and integrate the result over the inhomogeneouty region in QW. For convinience, we introduce the function
\begin{eqnarray}\label{A6}
u(y)=sgn(y)\frac{\phi(y)}{\Delta n}.
\end{eqnarray}
The function (\ref{A6}) is the ratio of the potential (\ref{A5}) to the concentration.
The extra concenration on Landau levels is
\begin{eqnarray}\label{A6}
\Delta n_{inh}(y)=n_L(y) - n_f
\end{eqnarray}
where $n_L(y)$ is inhomogeneous function of coordinate $y$. The additional requirements are $\left.\Delta n_{inh}(y)\right|_{y\leq0}=0$ and $\left.\exp(-oy)\Delta_{inh}(y)\right|_{y\to\infty}=0$ ($o$ is any positive real value). Then the averaged trough QW electrostatic potential $\phi_{inh}(y)$ could be calculated with the next formula
\begin{eqnarray}\label{A7}
\phi_{inh}(y)=\int_0^{y}\frac{\partial \Delta n_{inh}(y')}{\partial y'}u(y-y')dy', \qquad(y\ge0).
\end{eqnarray}
After integration by part one also could get more convinient formula:
\begin{eqnarray}\label{A8}
\phi_{inh}(y)=-\int_0^{y}\Delta n_{inh}(y')\frac{\partial u(y-y')}{\partial y'}dy', \qquad(y\ge0),
\end{eqnarray}
where we have used $u(0)=0$.

\appendix
\section*{Appendix B}
\setcounter{section}{2}
The effect of the ultrafast magnetic field dynamics on the carrier concentration is described below. We consider the strong inequality $\gamma\ll\omega_c^m$ to be true. This means that there is a time interval when the wave function $\Psi_n(t,x,y)$ is obtained by the substitution $\omega_c = \omega_c(t)$ in the stationary solution of the Schrodinger equation $\Psi_n^{(st.)}(x, y)$ (see [22 ] p. 424, the reduction to the 2D case is obvious) is a good zero approximation:
\begin{eqnarray}\label{B1}
\Psi_n(t,y,z) \simeq \exp\left[\int_0^t\hat{H}_0(t')dt'\right]\left.\Psi^{st.}_n(y,z)\right|_{\omega_c=\omega_c(t)}
\end{eqnarray}
Substituting Eq. (\ref{B1}) into the time-dependent Schrödinger Equation gives the equation with the term that could be considered as a perturbation:
\begin{eqnarray}\label{B2}
i\hbar\frac{\partial\Psi_n(t,y,z)}{\partial t} = \left[\hat{H}_0(t),\Psi_n(t,y,z)\right] +\left[\hat{V}(t),\Psi_n(t,y,z)\right],
\end{eqnarray}
where 
\begin{eqnarray}\label{B3}
\left[\hat{V}(t),\Psi_n(t,y,z)\right] = i\hbar\frac{\partial \omega_c}{\partial t}\frac{\partial\Psi_n(t,y,z)}{\partial \omega_c},
\end{eqnarray}
After going to the creation $\hat{a}^{+}_{p x,t}$ and annihilation $\hat{a}_{p_x,t}$ operators and excluding non-hermitian part one obtains
\begin{eqnarray}\label{B4}
\hat{V}_h(t)=\frac 12\left[\hat{V}(t) + h.c.\right]= \nonumber\\i\hbar\frac{\partial \ln|\omega_c(t)|}{\partial t}\left[\frac{\hat{a}_{p_x,t}{}^2-\hat{a}^{+}_{p_x,t}{}^2}{4}+\frac{\hat{p}_x}{\sqrt{2m\hbar\omega_c(t)}}\left(\hat{a}_{p_x,t}-\hat{a}^{+}_{p_x,t}\right)\right],
\end{eqnarray}
The Liouville equation for the system could be written as follows:
\begin{eqnarray}\label{B5}
i\hbar\frac{\partial\hat{\eta}(t)}{\partial t} = \left[\hat{H}_0(t),\hat{\eta}(t)\right] +\left[\hat{V}_h(t),\hat{\delta\eta}(t)\right],\nonumber\\
i\hbar\frac{\partial\hat{\delta\eta}(t)}{\partial t} = \left[\hat{H}_0(t),\hat{\delta\eta}(t)\right] +\left[\hat{V}_h(t),\hat{\eta}(t)\right];
\end{eqnarray}
where $\hat{\eta}(t)$ and $\hat{\delta\eta}(t)$ are the diagonal and off-diagonal elements of the density matrix. The first approximations for the solutions are
\begin{eqnarray}\label{B6}
\hat{\delta\eta}(t) \simeq -\frac{i}{\hbar}\int_{-\infty}^tdt'\exp\left[i\varphi(t,t')\hat{n}(t')\right]\left[\hat{V}(t'), \hat{\eta}_0(t')\right]\exp\left[-i\varphi(t,t')\hat{n}(t')\right],\nonumber\\
\hat{\eta}(t) \simeq \hat{\eta}_0(t)-\frac{i}{\hbar} \int_{-\infty}^tdt'\exp\left[i\varphi(t,t')\hat{n}(t')\right]\left[\hat{V}(t'), \hat{\delta\eta}(t')\right]\exp\left[-i\varphi(t,t')\hat{n}(t')\right];\nonumber\\
\end{eqnarray}
where $\hat{\eta}_{0}(t)$ is the densiy matrix built on states (\ref{B1}),  $\hat{n}(t)=\hat{a}^{+}_{p_x,t}\hat{a}_{p_x,t}$ and the phase $\varphi(t,t')$ is defined by (\ref{phase}). The population of the Landau levels is defenited by the expression:
\begin{eqnarray}\label{B7}
f_{n,\bar{y}}(t) = \left< n,t,\bar{y}|\hat{\eta}(t)|n,t,\bar{y}\right>,\quad f_{0n,\bar{y}}(t) = \left< n,t,\bar{y}|\hat{\eta}_0(t)|n,t,\bar{y}\right>;
\end{eqnarray}
where $\bar{y}(t) = p_x/[m\omega_c(t)]$. Using (\ref{B6}) and (\ref{B7}) we get the expression:
\begin{eqnarray}\label{B8}
\frac{\partial f_{n,\bar{y}}(t)}{\partial t} =\frac{\partial f_{0n,\bar{y}}(t)}{\partial t}+\frac{\partial \ln|\omega_c(t)|}{\partial t}\nonumber\\ \times\int_{-\infty}^tdt'\frac{\partial \ln|\omega_c(t')|}{\partial t'}\sum_{k=-2}^2\nu_{n,n+k}(t,t')\left[f_{n+k,\bar{y}}(t')-f_{n,\bar{y}}(t')\right],
\end{eqnarray}
the koeficients $\nu_{n,n'}$ are defined as follows:
\begin{eqnarray}\label{B9}
\nu_{n,n+k,\bar{y}}(t,t') \nonumber\\=\left\{
\begin{array}{lll}
\frac 18(n+k/2)(n+1+k/2)\cos 2\varphi(t,t'),\quad  n+k \ge 0, ~ k=\pm 2;\\
n\frac{\omega_c(t')}{\omega_c(t)}\left[\frac{\bar{y}(t')}{l_c(t')}\right]^2\cos \varphi(t,t'), \qquad n+k \ge 0,~ k=\pm 1;\\
0,\qquad n+k <0.
\end{array}
\right.
\end{eqnarray}
whereas $l_c(t)=\sqrt{\hbar/[m\omega_c(t)]}$. 
Summing over all Landau levels one cold obtain that the second term in (\ref{B7}) does not change the concentration
\begin{eqnarray}\label{B10}
\sum_{n=0}^{\infty}\frac{\partial f_{n,\bar{y}}(t)}{\partial t} =\sum_{n=0}^{\infty}\frac{\partial f_{0n,\bar{y}}(t)}{\partial t}
\end{eqnarray}
Since, while maintaining the same accuracy, in the subintegral expression of (\ref{B8}) we can replace $f_n(t')=f_n(t)+\sum_{n=0}^{\infty}\frac{\partial f_{0n}(t)}{\partial t}\frac{(t-t')^n}{n!}$, then we can estimate the quantum lifetime as:
\begin{eqnarray}\label{B11}
 \tau_{n}^{-1}(t,\bar{y})\approx\frac{\partial\ln|\omega_c(t)|}{\partial t}\int_{-\infty}^tdt'\frac{\partial \ln|\omega_c(t')|}{\partial t'}\sum_{k=-2}^2|\nu_{n,n+k}(t,t',\bar{y})|,
\end{eqnarray}
where we can use absolute values of (\ref{B9}) because the summation performed for the symmetric range.

\end{document}